# Magnetization reversal of a ferromagnetic Pt/Co/Pt film by helicity dependent absorption of visible to near-infrared laser pulses

Kihiro T. Yamada,[1,2,*,†] Carl S. Davies,[3,‡,†] Fuyuki Ando,[4] Tian Li,[4] Teruo Ono,[4,5,6]
Theo Rasing,[1] Alexey V. Kimel[1] and Andrei Kirilyuk[1,3]

*[1]Radboud University, Institute for Molecules and Materials, Nijmegen 6525 AJ, The Netherlands*

*[2]Department of Physics, Institute of Science Tokyo, Tokyo 152-8551, Japan.*

*[3]FELIX Laboratory, Radboud University, Toernooiveld 7, Nijmegen 6525 ED, The Netherlands.*

*[4]Institute for Chemical Research, Kyoto University, Uji 611-0011, Japan.*

*[5]Center for Spintronics Research Network, Institute for Chemical Research, Kyoto University, Uji, 611-0011, Japan*

*[6]Center for Spintronics Research Network (CSRN), Graduate School of Engineering Science, Osaka University, Toyonaka 560-8531, Japan.*

[†]These authors contributed equally.

Corresponding authors: [*]yamada.k.9463@m.isct.ac.jp; [‡]c.davies@science.ru.nl

## Abstract

The practical difficulty in distinguishing the impact of magnetic circular dichroism and the inverse Faraday effect fuels intense debates over which mechanism predominantly drives the process of helicity dependent all-optical switching of magnetization in ferromagnets. Here, we quantitatively measure the efficiency of the switching process in a Pt/Co/Pt multilayer stack using visible- to near-infrared optical pulses. We find that the switching efficiency increases by a factor of 8.6 upon increasing the pumping wavelength from 0.5 μm to 1.1 μm, becoming 100% efficient at even longer wavelengths up to 2.0 μm. Our experimental results can be successfully explained by the phenomenon of magnetic circular dichroism, making a significant step towards resolving the long-standing controversy over the origin of the all-optical process of magnetization reversal in ferromagnets.

The discovery of demagnetization on a sub-picosecond time scale by a femtosecond (fs) laser pulse gave birth to the new research field of ultrafast magnetism [1]. Furthermore, the use of circularly-polarized pulses allows magnetization dynamics to be selectively excited according to the optical helicity [2-4]. Helicity-dependent control of magnetization precession was first demonstrated in the dielectric $DyFeO_3$ by Kimel *et al.* [2], which was attributed to the Inverse Faraday effect (IFE) [3]. Subsequently, similar effects have also been observed in magnetic semiconductors [4] and metals [5-6]. For some magnetic metals, magnetization can be switched by fs circularly-polarized laser pulses in the absence of an external magnetic field. This phenomenon is referred to as all-optical helicity-dependent switching (AO-HDS). AO-HDS was first demonstrated in ferrimagnetic GdFeCo [7] and later in ferromagnetic metals, including multilayered Co/Pt [8-15] and Co/Ni [8,10] stacks and granular FePt [16-17]. All-optical magnetization switching in GdFeCo films is driven by fast and efficient heating of electrons by the fs laser pulse [18] that brings the ferrimagnetic medium to a strongly non-equilibrium state [19]. The helicity dependence of the switching is derived from magnetic circular dichroism (MCD), which leads to different laser absorption levels that depend on the helicity [20], but the switching, in general, relies only on the fast heating. In ferromagnetic metals, in contrast, it is evident that circularly polarized pulses are genuinely indispensable for AO-HDS. However, when discussing the dominant mechanism that causes the switching in such systems, one always encounters the outstanding question: do non-thermal or thermal effects drive the AO-HDS?

On the one hand, the IFE is well known to be non-dissipative [3,21]. A circularly polarized fs pulse induces a magnetization in magnetic media via impulsive stimulated Raman processes [22], which is not accompanied by absorbing photons. Microscopic three-temperature models [23] and an atomistic spin model [24] all show that the magnetic polarity of a ferromagnetic metal can indeed be switched by the IFE.

On the other hand, the absorption of a fs laser pulse by a magnetic metal quasi-instantaneously elevates the metal's temperature on a non-equilibrium timescale, resulting in ultrafast demagnetization [1]. Because the optical absorption of a ferromagnetic metal varies according to the helicity via the MCD, helicity dependent laser absorption may be relevant to AO-HDS [24-25]. When magnetic domains of opposite polarity are illuminated by a circularly polarized pulse, their free energies decrease depending on the combination of optical helicity and magnetic orientation. As a result of repeating this process, magnetic domains with lower free energies should expand and eventually form a single magnetic domain, the final orientation of which depends on the helicity of the circularly polarized pulse**.**

Through testing how the pulse-width [11] and starting-temperature [13] affect AO-HDS in Pt/Co/Pt heterostructures, previous works have suggested that the MCD effect drives the AO-HDS. For the case of AOS of magnetization in ferrimagnetic GdFeCo alloys, its spectral dependence was used to exclude the possible contribution of the IFE [20]. Investigating the spectral dependence of AO-HDS could similarly resolve the dominant driving mechanism of AO-HDS in Pt/Co/Pt multilayer stacks.

In this Letter, we investigate multi-pulse AO-HDS of a ferromagnetic Pt/Co/Pt system from the visible to near-infrared spectral ranges to elucidate whether AO-HDS is dominated by thermal or non-thermal effects. We experimentally find that the switching efficiency – defined as the amount of magnetization that is selectively switched by circularly-polarized light – is enhanced by a factor of 8.6 upon increasing the optical wavelength from 0.5 μm to 1.2 μm. A phenomenological model of the inverse Faraday effect cannot explain this improved switching efficiency. Instead, we find that the effective magnetic field originating from the helicity dependent laser absorption increases by a factor of 32 in this spectral range, convincingly demonstrating that our results can be quantitatively described as a consequence of MCD.

Perpendicularly-magnetized Pt/Co/Pt structures represent the most standard systems for studies of AO-HDS of ferromagnets [8-15]. We therefore used similar materials, depositing Ta (1.0 nm)/MgO (2.0 nm)/Pt (3.0 nm)/Co (0.6 nm)/Pt (3.0 nm)/Ta (4.0 nm) on a synthetic glass substrate by magnetron sputtering. We summarize in Fig. 1 the Faraday rotation $\theta_F$ and ellipticity $\eta_F$ of the Pt/Co/Pt stack in the range of wavelengths $\lambda$ between 0.5 μm and 1.2 μm, measured using a monochromator in combination with a photo-elastic modulator [26]. Figure 1 shows that $\theta_F$ and $\eta_F$ increase, within this spectral range, by factors of ~3.0 and ~2.0 respectively. We also determined the magneto-optical parts of the refractive indices of the Co and Pt layers using a transfer matrix method (see Ref. [27] and Supplemental Note 1 [28] for details of this method). We thus calculate the MCD defined as $\mathrm{MCD} = (A^- - A^+) / [(1/2)\,(A^+ + A^-)]$, where $A^{+(-)}$ is the total absorption of right- (left-) circularly-polarized light by the up-magnetized film. The spectral dependence of MCD is also plotted in Fig. 1, showing that MCD increases by a factor of ~4.2 with increasing $\lambda$ in the considered spectral range.

To excite the Pt/Co/Pt system, we used optical pulses ($\lambda$ = 800 nm) delivered by an amplified Ti: Sa laser system at a repetition rate of 1 kHz. For $\lambda$ = 800 nm, the pulse width was characterized using an autocorrelator to be 60 fs. By pumping an optical parametric amplifier (OPA), the central wavelength can be adjusted in the range $\lambda$ = 0.5 – 2.0 μm. To obtain circular polarization across this broad spectral range, we used suitable quarter-wave plates AQWP05M-600 and AQWP05M-980 (Thorlabs Inc.) for $\lambda$ = 0.5 – 0.8 μm and 0.95 – 1.1 μm, respectively, and PO-TWP-L4-25-IR (ALPHALAS GmbH) for $\lambda$ = 1.2 – 2.0 μm. We found the ellipticity of the pump laser pulse induced by these achromatic quarter-wave plates, at each wavelength, to be consistently larger than 0.9. The train of circularly-polarized optical pulses was focused on the surface of the uniformly-magnetized Pt/Co/Pt sample. We characterized the Gaussian spot size on the sample at each wavelength by Liu's method [20,29]. The incident fluence was determined by using the extracted spot size and the laser power measured in front of the sample. To evaluate the switching efficiency of AO-HDS, we perform sweeping experiments whereby the sample is mounted on a motorized stage, enabling the laser pulses to be swept 100 μm across the sample at a constant speed of 10 μm/s. We used the magneto-optical Faraday effect to directly visualize the magnetization after exposure to the multiple optical pulses. Linearly-polarized white light illuminates the sample and is collected by a ×20 objective lens. Depending on whether the magnetization of the sample is parallel or antiparallel to the wave vector of the transmitted light, the latter's polarization rotates in different senses. An analyzer, therefore, enables an image of magnetization to be detected using a charge-coupled device camera.

Figure 2(a) displays typical snapshots recorded after sweeping the Pt/Co/Pt sample with pulses at $\lambda$ = 0.5 μm, 0.8 μm, 1.1 μm and 2.0 μm for the four cases ($M^{\uparrow}$, $\sigma^{+}$), ($M^{\uparrow}$, $\sigma^{-}$), ($M^{\downarrow}$, $\sigma^{+}$) and ($M^{\downarrow}$, $\sigma^{c}$) respectively. Here, we use the notation ($M^{\uparrow\downarrow}$, $\sigma^{\pm}$) to indicate that the sample had initial magnetization pointing up $M^{\uparrow}$ or down $M^{\downarrow}$ and the impinging optical pulses were circularly-polarized with right-handed $\sigma^{+}$ or left-handed $\sigma^{-}$ helicity. Here, the electric field vectors of $\sigma^{+}$ and $\sigma^{-}$ light projected on a screen rotate clockwise and anticlockwise, respectively, when standing against the light source. Figure 2(a) clearly shows that AO-HDS appears to become more efficient with increasing $\lambda$. To quantify the spectral dependence of the AO-HDS, we estimate the net switched magnetization $<M>$ as follows. First, we averaged the intensities in a central rectangular area spanning 80 μm × 20 μm (364 × 91 pixels) on the track of the laser spot, as shown in Fig. 2(a).

Second, <*M*> was estimated by normalizing the average with a reference image recorded for the sample in a uniformly-magnetized state. Figure 2(b) shows <*M*> as a function of the laser incident fluence *F* for $\lambda$ = 0.5 μm, 0.8 μm, 1.1 μm, and 2.0 μm. The helicity dependence of the switching becomes increasingly stronger with increasing $\lambda$. In particular, the helicity dependence emerges when the fluence exceeds the demagnetization threshold [Fig. S2], where switched domains start to appear at the center of the spot because the laser fluence is high enough to excite the magnetization beyond the Curie temperature. As previously reported [8], AO-HDS demands heating of the system close to the Curie temperature. Figure 3(a) plots the extracted switching efficiency as $\varepsilon = \langle M \rangle \; [(M^{\uparrow}, \sigma^{+}) - (M^{\uparrow}, \sigma^{-}) + (M^{\downarrow}, \sigma^{+}) - (M^{\downarrow}, \sigma^{-})]/4$ versus $\lambda$, where we average the net magnetization <*M*> obtained for all fluences larger than that defining the demagnetization threshold. The spectral dependence of $\varepsilon$ represents our key experimental finding and will be used later for quantitatively discussing which mechanism - the IFE or the MCD effect - principally drives the AO-HDS. The switching efficiency $\varepsilon$ increases by a factor of ~8.6 from $\lambda$ = 0.5 μm to $\lambda$ = 1.1 μm; $\varepsilon$ reaches 100 % (i.e., full switching) when $\lambda \geq$ 1.2 μm. Simultaneously, we find that the minimum incident optical fluence required for AO-HDS decreases quite substantially as $\lambda$ increases, that is, the process becomes more energy-efficient for longer wavelengths (see Fig. 2 (b)).

The pulse width is known to have a significant influence on AO-HDS. Extending the pulse durations to several picoseconds when optically irradiating multilayered $[Pt/Co]_3$ stacks, for example, significantly promotes the switching efficiency compared to using 60-fs pulses [11]. It is therefore important to assess the pulse durations delivered across the broad range of wavelengths supplied by the OPA. In lieu of suitably broadband standard detection methods, we indirectly assessed the pulse width of the OPA output using single-shot all-optical switching in GdFeCo [30,31] (see the Supplemental Note 3 [28]). Figure S4 shows that the pulse width is nearly constant at ~100 fs for all wavelengths tested, except for an anomaly around 0.55-0.75 μm where it rises to 200-350 fs. While our employed method for measuring the pulse-duration is not as precise as standard techniques (such as frequency-resolved optical gating [32]), the obtained results plausibly show that the observed variation of $\varepsilon$ cannot be explained by an elongation of the pulse width with increasing $\lambda$.

We first evaluate the spectral dependence of AO-HDS originating from the helicity dependent laser absorption via MCD. In general, this model [9,11,14] explains AO-HDS in terms of a two-step process. In the first step, the thermal load delivered by an optical pulse induces demagnetization in the central area of the Gaussian spot, creating many domain-walls (DWs) within a multi-domain state. In the second step, subsequent circularly-polarized optical pulses drive DW motion. The direction of the latter is defined by the MCD, since the circularly-polarized optical pulse is preferentially absorbed by one of the magnetic domains compared to the other. Because the 'hotter' domain has smaller free energy than the cooler domain, the DW moves opposite to the direction of the heat current [11]. Thus, the temperature gradient induced by the difference in optical absorption (originating from MCD) leads to DW motion that ultimately switches the net magnetization.

To model this process, a DW excited by a circularly polarized pulse experiences an effective magnetic field $B_{\text{eff}}^{\text{MCD}}$ created by a one-dimensional thermal gradient $\partial T/\partial x$ through the MCD [33]:

$$B_{\text{eff}}^{\text{MCD}} = -\frac{2}{Ml}\frac{\partial A}{\partial T}\frac{\partial T}{\partial x}, \tag{1}$$

where $M$, $A$, and $l$ denote the saturation magnetization, exchange stiffness, and domain-wall width, respectively. Assuming that the irradiated region is heated from room temperature ($T_0$ = 300 K) towards the Curie temperature ($T_C$ = 470 K) [34], the one-dimensional thermal gradient acting across the DW is $\partial T/\partial x$ = MCD ($T_C - T_0$) / $l$. Note that the thermal gradient is on the order of 0.01-0.1 K/nm, which is four orders of magnitude larger than that required to move the DW of a magnetic garnet [35]. Using the temperature dependences of $M$, $A$, and $l$ given in Ref. [36], we estimate the $B_{\mathrm{eff}}^{\mathrm{MCD}}$ from: $M(T) = M(0)[1 - (T / T_C)^{2.37}]^{0.34}$, $A(T) = A(0)[M(T) / M(0)]^{1.82}$, $l(T) = l(0)[M(T) / M(0)]^{-0.59}$. We also simply assume the temperature dependence of the MCD as MCD($T$) = MCD(0)[$M(T)$ / $M(0)$]. The values at $T$ = 0 K are estimated by using $M(300) = 1.13\times10^{6}$ A/m, $A(300) = 1.6\times10^{-11}$ J/m, and $l(300)$ = 6.6 nm following Ref. [34]. The results are plotted in Fig. 3(b) and show a monotonic increase of $B_{\mathrm{eff}}^{\mathrm{MCD}}$ with wavelength, similar to the experimentally observed increase of the efficiency in that spectral range [Fig. 3(a)].

We next estimate the $B_{\mathrm{eff}}^{\mathrm{IFE}}$ produced by the IFE using a phenomenological approach [21]. The IFE induces a magnetization along the direction given by the cross product of the light's electrical field and its complex conjugation, $\mathbf{E} \times \mathbf{E}^*$. The resultant magnetization, which is qualitatively equivalent to an effective magnetic field, is therefore nominally capable of switching the magnetization direction. In an isotropic magnetic medium, the strength of the effective magnetic field of the IFE [21] is given by

$$B_{\mathrm{eff}}^{\mathrm{IFE}} = \frac{\theta_F \lambda n F}{\pi M c \tau d}, \qquad (2)$$

where $c$, $\tau$, $d$, and $F$ denote the speed of light, the pulse width, the stack thickness, and the optical fluence, respectively. For the estimation, we fixed the pulse width at 60 fs for all wavelengths and used the thickness-weighted average of the real parts of the refractive indices of the Co and Pt layers [28]. Here, we considered the sum of the fluences, at the threshold of demagnetization, between an optical pulse propagating downward from the MgO/Pt interface and one propagating upward from the Pt/Ta interface. Although the handedness of a circularly polarized pulse changes by reflection, the IFE of the reflected pulse has the same sign as that of the incident pulse because of the inversion of the $k$ vector. Figure 3(b) indicates that the calculated effective magnetic field induced by the IFE has a rather different spectral dependence. Although the $B_{\mathrm{eff}}^{\mathrm{IFE}}$ also initially rises, it has a clear maximum at $\lambda$ = 0.8 μm, whereafter it decreases to almost half from the maximum at $\lambda$ = 1.1 μm. This is in stark contrast with the clear experimental observation of $\varepsilon$ scaling monotonically with increasing $\lambda$ [Fig. 3(a)].

In the spectral range of 0.5 μm – 1.1 μm, the switching efficiency $\varepsilon$ increases monotonically by a factor of ~8.6 [Fig. 3(a)]. In the same spectral range, the $B_{\mathrm{eff}}^{\mathrm{MCD}}$ also increases monotonically (by a factor of ~32) while the spectral dependence of $B_{\mathrm{eff}}^{\mathrm{IFE}}$ does not accord with that of $\varepsilon$ [Fig. 3(b)]. Thus, the dramatic improvement of the AO-HDS with increasing $\lambda$ can quantitatively be explained by the spectral dependence of the MCD effect but not by that of the IFE**.** Therefore, we conclude that the MCD effect predominantly drives the AO-HDS.

To ensure that our conclusion is not limited to a single Pt/Co/Pt stack, we also investigated the spectral dependence of AO-HDS found in a Pt (3 nm)/Co (0.8 nm)/Pt (3 nm) multilayer using the same procedures (see Supplemental Note 4 [28]). We observed a similar monotonic improvement of AO-HDS with increasing $\lambda$ for this thicker multilayer, which is explained by the monotonically increasing MCD effect. Moreover, we studied the AO-HDS in a [Pt (2 nm)/Co (0.8 nm)]$_2$/Pt (2 nm) heterostructure. Again, for the multilayer with different lamination, we observed a similar enhancement of AO-HDS with increasing $\lambda$ (see Supplemental Note 4 [28]). These results, obtained for different nanolayer-thicknesses and laminations, indicate the robustness of our conclusion – namely, that the MCD effect dominates AO-HDS for ferromagnets in the visible to near-infrared spectral ranges.

Note that the duration of $B_{\mathrm{eff}}$ is also important for magnetization switching. On the one hand, the impulsive stimulated Raman process associated with the IFE is only active while the optical pulse is present in the system, and so the duration of $B_{\mathrm{eff}}^{\mathrm{IFE}}$ is equivalent to the duration of the employed fs pulse. On the other hand, the duration of $B_{\mathrm{eff}}^{\mathrm{MCD}}$ will be much longer, because the iso-thermalization of two immediately-adjacent magnetic domains will proceed across timescales on the order of hundreds of picoseconds or more. Therefore, although the peak amplitude of the $B_{\mathrm{eff}}^{\mathrm{MCD}}$ is smaller, the temporal integration of its effect will be much larger than that obtained from $B_{\mathrm{eff}}^{\mathrm{IFE}}$.

We also note the limitation of the phenomenological model of the inverse Faraday effect: the Faraday rotation does not necessarily agree with the inverse Faraday effect for absorbing magnetic media [37]. Ab-initio calculations may be useful to more accurately discuss the contribution of the IFE to the AO-HDS. It is anticipated from Eq. (2) and the ab-initio calculations from Ref. [38] that $B_{\mathrm{eff}}^{\mathrm{IFE}}$ diverges for increasing $\lambda$ and a constant fluence. Investigating AO-HDS with infrared pulses in a free-electron laser facility [39,40] is a future work planned to gain more insight into the possibility of using IFE to achieve AO-HDS. Although we have tested three different Pt/Co/Pt multilayered stacks, AO-HDS is observed also in other thin ferromagnetic [8,10] as well as ferrimagnetic [41,42] metals. We anticipate that a similar methodological study of a much larger class of materials could ultimately lead to an even better generalized understanding of AO-HDS. In particular, a sign inversion of the Kerr ellipticity close to $\lambda$ = 500 nm was recently reported in a sperimagnetic alloy of TbCo [26]. Assuming that MCD drives the switching, pumping this sample at wavelengths above and below this threshold would presumably reverse the symmetry of AO-HDS.

To summarize, we have investigated the spectral dependence of AO-HDS in a Pt/Co/Pt multilayered stack, in the visible to the near-infrared spectral regime, to ascertain the dominant mechanism that drives the process. We have experimentally found that the multi-pulse AO-HDS effect becomes increasingly efficient upon increasing the excitation wavelength, reaching 100% at wavelengths greater than 1.2 μm. The monotonic enhancement of the switching efficiency follows that of the effective field associated with the helicity dependent laser absorption (MCD), while the inverse Faraday effect cannot explain the improved AO-HDS. Our results provide crucial and experimentally-grounded insight into the origin of AO-HDS in ferromagnetic metals.

## Acknowledgements

The authors thank S. Semin and C. Berkhout for technical support, and A. Tsukamoto for providing the GdFeCo sample. This research has received funding from de Nederlandse Organisatie voor Wetenschappelijk Onderzoek (NWO) and the European Union's Horizon 2020 research and innovation program under FET Open Grant Agreement No. 713481 (SPICE) and No. 737093 (FEMTOTERABYTE) and ERC grant agreement No.856538 (3D-MAGiC). This work was partially supported also by JSPS KAKENHI No. 22K14588, No. 24K00938, No. 15H05702, No. 20K22327, No.20H00332, No. 20H05665, No. 17J07326, and No. 18J22219, by the Collaborative Research Program of the Institute for Chemical Research, Kyoto University, and by the Spintronics Research Network of Japan (Spin-RNJ).

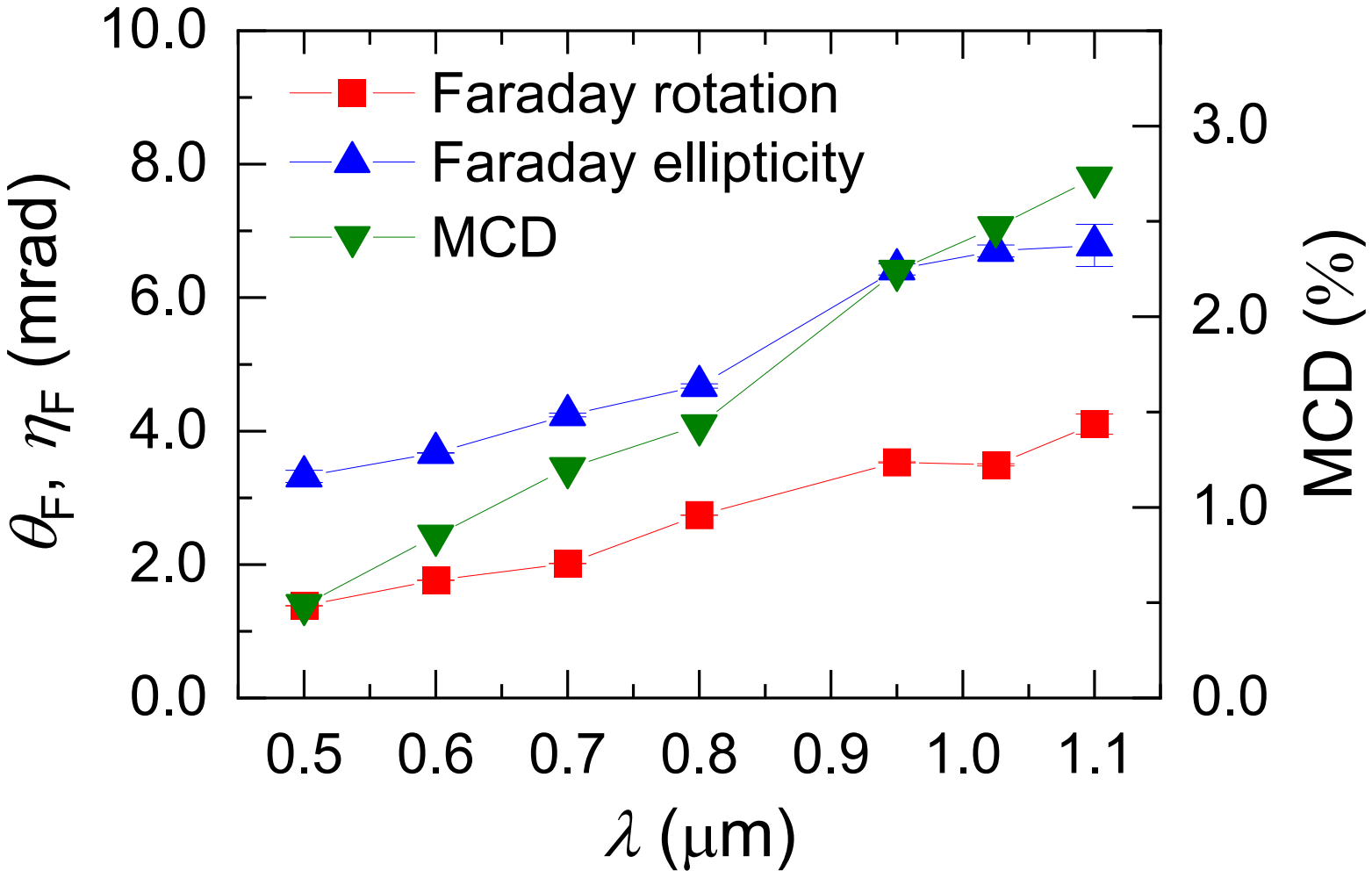

FIG. 1. The experimentally-measured spectral dependences of Faraday rotation $\theta_F$, Faraday ellipticity $\eta_F$ and magnetic circular dichroism MCD of the studied Pt/Co/Pt multilayered stack.

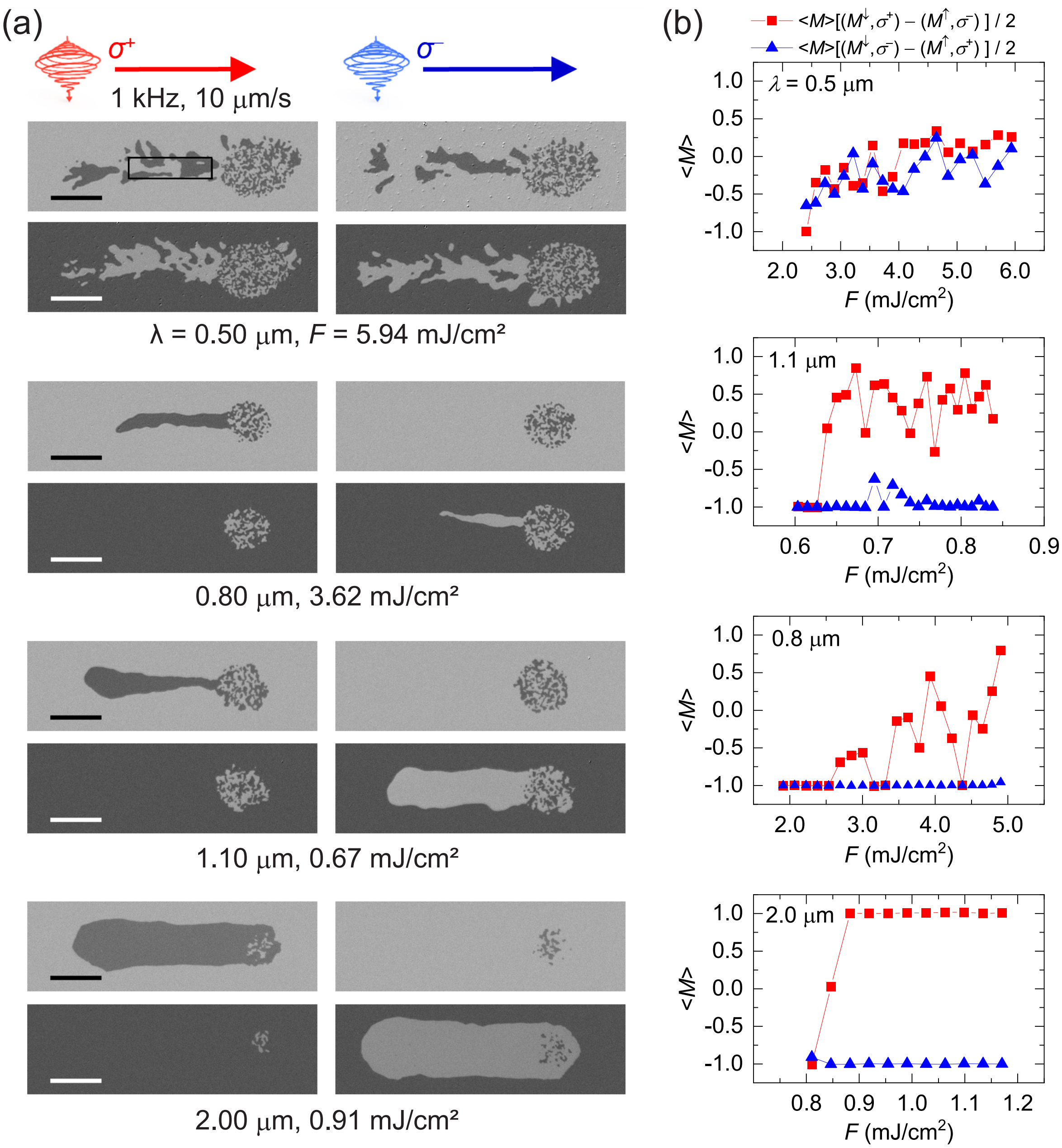

FIG. 2. (a) Typical results of multi-pulse all-optical helicity dependent switching (AO-HDS) for pumping wavelengths $\lambda$ of 0.5 μm, 0.8 μm, 1.1 μm, and 2.0 μm as indicated. Dark and bright contrasts correspond to up- and down-magnetized domains (pointing into or out of the plane of the page), respectively. Incident optical fluences ($F$) are shown below the snapshots. The scale bars correspond to 30 μm. We integrated the intensity in the rectangle of 20 μm × 80 μm to evaluate the net magnetization <*M*>. (b) <*M*> as a function of $F$ for $\lambda$ = 0.5 μm, 0.8 μm, 1.1 μm and 2.0 μm. The red (blue) points correspond to the case when the spin angular momentum of light is antiparallel (parallel) to the magnetization direction of the original state.

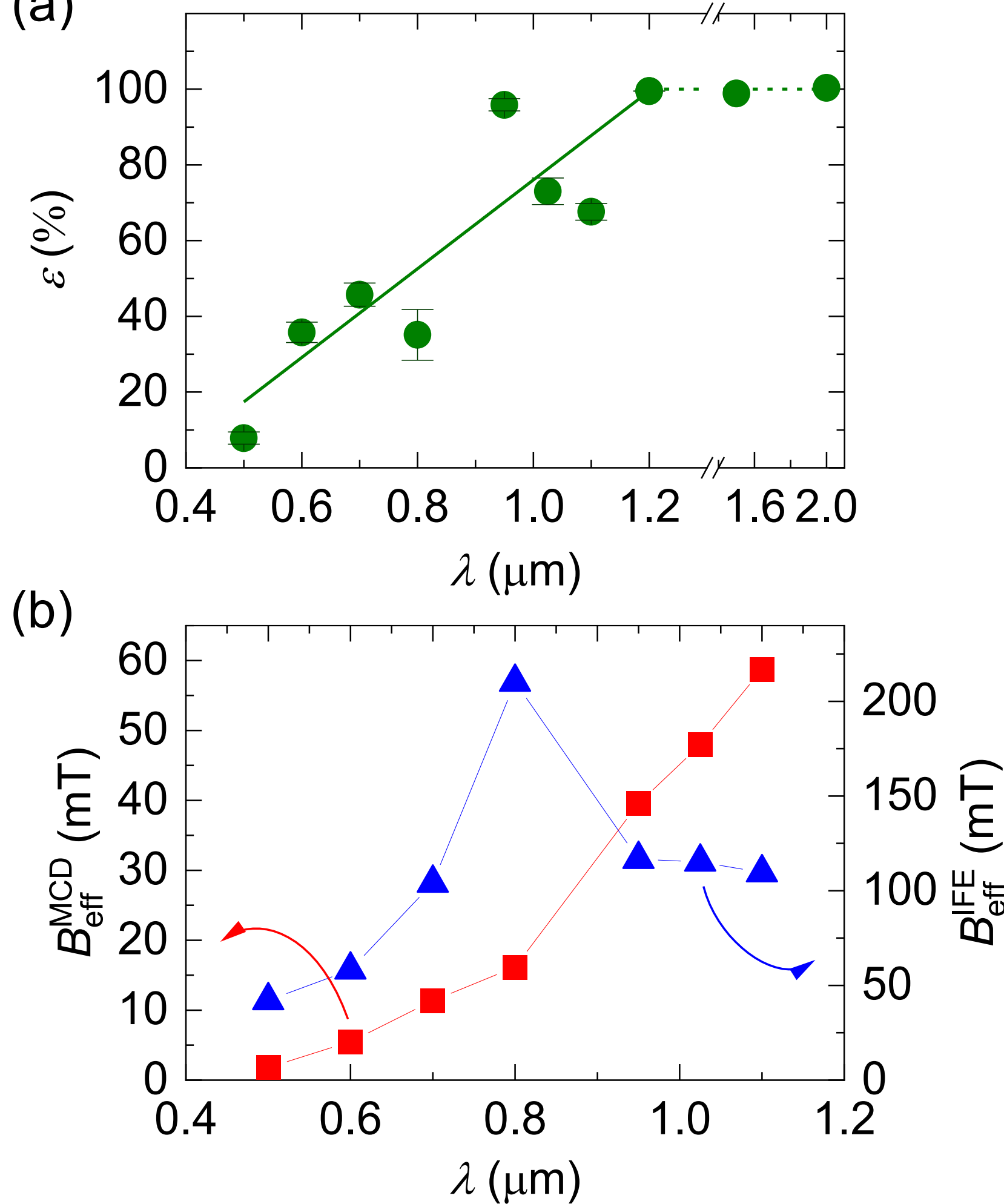

FIG. 3. (a) The spectral dependence of the switching efficiency $\varepsilon$, obtained by averaging the efficiencies measured for varying incident fluences above the threshold of demagnetization. The solid line is a linear fit for points between $\lambda = 0.5$ μm and 1.1 μm. (b) The spectral dependencies of the effective fields $B_{eff}$ associated with the inverse Faraday effect (blue triangles) and the MCD effect (red squares).

# Supplemental Material

## Magnetization reversal of a ferromagnetic Pt/Co/Pt film

## by helicity dependent absorption of visible to near-infrared laser pulses

Kihiro T. Yamada,[1,2,*,†] Carl S. Davies,[3,‡,†] Fuyuki Ando,[4] Tian Li,[4] Teruo Ono,[4,5,6]

Theo Rasing,[1] Alexey V. Kimel[1] and Andrei Kirilyuk[1,3]

[1]*Radboud University, Institute for Molecules and Materials, Nijmegen 6525 AJ, The Netherlands*

[2]*Department of Physics, Institute of Science Tokyo, Tokyo 152-8551, Japan.*

[3]*FELIX Laboratory, Radboud University, Toernooiveld 7, Nijmegen 6525 ED, The Netherlands.*

[4]*Institute for Chemical Research, Kyoto University, Uji 611-0011, Japan.*

[5]*Center for Spintronics Research Network, Institute for Chemical Research, Kyoto University, Uji, 611-0011, Japan*

[6]*Center for Spintronics Research Network (CSRN), Graduate School of Engineering Science, Osaka University, Toyonaka 560-8531, Japan.*

[†]These authors contributed equally.

Corresponding authors: [*]yamada.k.9463@m.isct.ac.jp; [‡]c.davies@science.ru.nl

**Supplemental Note 1: Estimation of the magneto-optical part of the refractive index based on a transfer-matrix formalism**

To calculate the amplitude of the light electric field at every interface, one must consider not only the reflection and transmission of light at every interface between two different layers but also the absorption of light. Therefore, we used a well-known transfer matrix formalism [S1] to obtain the electric fields inside our multilayer by considering the refractive indices, $\tilde{n} = n + jk$, of the employed layers (Table S1) at each wavelength $\lambda$. For normal incidence at a multilayer as depicted in Fig. S1, the Fresnel coefficients for the $m$-th interface are common for s- and p-polarized light:

$$r_{m,m+1} = \frac{n_m - n_{m+1}}{n_m + n_{m+1}}, \qquad t_{m,m+1} = \frac{n_m}{n_m + n_{m+1}}. \tag{1}$$

Using the Fresnel coefficients, the electric fields on the left-hand and right-hand sides of the $m$-th interface are connected with an interface matrix $I_\mathrm{m}$ as

$$\begin{pmatrix} E_m^{L\rightarrow} \\ E_m^{L\leftarrow} \end{pmatrix} = \frac{1}{t_{m,m+1}} \begin{pmatrix} 1 & r_{m+1,m} \\ r_{m,m+1} & 1 \end{pmatrix} \begin{pmatrix} E_m^{R\rightarrow} \\ E_m^{R\leftarrow} \end{pmatrix} \equiv I_m \begin{pmatrix} E_m^{R\rightarrow} \\ E_m^{R\leftarrow} \end{pmatrix}. \tag{2}$$

The amplitudes and phases of the electric fields change as they travel across the $m$-th layer with a thickness $d_m$ and a refractive index $\widetilde{n_m}$. The electric fields at the right-hand side of the $m$-th interface are connected with the ones on the left-hand side of the ($m$ + 1)-th interface via a transport matrix $T_\mathrm{m}$ according to

$$\begin{pmatrix} E_m^{R\rightarrow} \\ E_m^{R\leftarrow} \end{pmatrix} = \begin{pmatrix} \exp(-2\pi j \tilde{n}_m d_m/\lambda) & 0 \\ 0 & \exp(2\pi j \tilde{n}_m d_m/\lambda) \end{pmatrix} \begin{pmatrix} E_{m+1}^{L\rightarrow} \\ E_{m+1}^{L\leftarrow} \end{pmatrix} \equiv T_m \begin{pmatrix} E_{m+1}^{L\rightarrow} \\ E_{m+1}^{L\leftarrow} \end{pmatrix}. \tag{3}$$

Therefore, the electric fields on the left-hand side of the $m$-th interface are expressed in the electric fields at the left-hand side of the ($m$ + 1)-th interfaces as

$$\begin{pmatrix} E_m^{L\rightarrow} \\ E_m^{L\leftarrow} \end{pmatrix} = I_m T_m \begin{pmatrix} E_{m+1}^{L\rightarrow} \\ E_{m+1}^{L\leftarrow} \end{pmatrix}. \tag{4}$$

By repeating this process for all the interfaces and layers, one can obtain the relationship between the electric field amplitude of the incident (*i*), the reflected (*r*), and the transmitted (*t*) light waves:

$$\begin{pmatrix} i \\ r \end{pmatrix} = I_1 T_1 \cdots I_m T_m \cdots I_n T_n I_{n+1} \begin{pmatrix} T \\ 0 \end{pmatrix} \equiv \begin{pmatrix} A_{11} & A_{12} \\ A_{21} & A_{22} \end{pmatrix} \begin{pmatrix} t \\ 0 \end{pmatrix}. \tag{5}$$

Solving this gives important relationships:

$$r = \left(\frac{A_{21}}{A_{11}}\right) i, \qquad t = \left(\frac{1}{A_{11}}\right) i. \tag{6}$$

From energy conservation, the absorbed energy is

$$A = i^* i - r^* r - t^* t, \tag{7}$$

where '*' denotes the complex conjugation. Also, the electric field amplitudes at a specific interface can be calculated with the interface and transport matrices; for instance, the electric field amplitudes at the *m*-th interface are

$$\begin{pmatrix} E_m^{R\rightarrow} \\ E_m^{R\leftarrow} \end{pmatrix} = (I_1 T_1 \cdots I_{m-1} T_{m-1})^{-1} \begin{pmatrix} i \\ A_{21} i / A_{11} \end{pmatrix}. \tag{8}$$

The reflection changes the wave vector and the handedness of circularly polarized light. We note that applying the transfer matrix method to the magnetic layers requires consideration of not only the wave vector and handedness of circularly polarized light but also the magnetic orientations.

We estimated the refractive indices of the Co and Pt layers using the transfer matrix formalism for normal incidence. Throughout the estimation, for simplicity, we assumed that the magneto-optical parts of $\tilde{n}$ (i.e., $n^+ - n^-$ and $k^+ - k^-$) are identical in the Pt/Co/Pt layer. The $n^{+(-)}$ and $k^{+(-)}$ denote the real and imaginary parts of the refractive index for right(left)-handed circularly polarized light, respectively. The assumption that the magneto-optical effect arises only from the Co layer gave unrealistically large values of $n^+$, $n^-$, $k^+$, and $k^-$. The Faraday rotation $\theta_F$ is expressed as

$$\theta_F = \frac{\pi d}{\lambda}(n^+ - n^-), \tag{9}$$

where $d$ is the total thickness of the Pt/Co/Pt multilayer. We determined the $n^+$ and $n^-$ directly from the experimental value of $\theta_F$. The Faraday ellipticity $\eta_F$ is expressed as

$$\eta_F = \tan^{-1}\left(\frac{|E^+| - |E^-|}{|E^+| + |E^-|}\right), \tag{10}$$

where $E^{+(-)}$ is the electric field of the transmitted right (left)-handed circularly polarized light. Using the transfer-matrix method, we determined values of $k^+$ and $k^-$ that match the experimental value of $\eta_F$ for the up-magnetized case. Table S1 summarizes the estimated values of $\tilde{n}$ of Co and Pt layers when the magnetization is antiparallel to the light wave vector. We moreover determined the magnetic circular dichroism [Fig. 1(c)] with the determined refractive indices based on Eq. (8).

**Supplemental Note 2: Demagnetization threshold**

In the main text, we defined the demagnetization threshold as the optical fluence at which switched domains start appearing at the center of the laser spot as in Fig. S2(a). Figure S2(b) shows the demagnetization threshold and the absorbed energy as a function of $\lambda$. While the absorbed energy at the demagnetization threshold is almost constant up to 800 nm, it decreases by a factor of 4.2 with further increasing of $\lambda$.

**Supplemental Note 3: Assessing the wavelength-dependent pulse duration**

The process of AO-HDS is well-known to depend on the duration of the optical pulses. Ref. [S2] showed, for example, that 4-ps-long circularly polarized pulses ($\lambda = 800$ nm) are much more efficient in switching magnetization compared to fs pulses. It is therefore important for us to evaluate the duration of the visible and near-infrared pulses delivered by the OPA. In lieu of suitable broadband autocorrelators, we indirectly estimate the pulse width by measuring the spectral and thermal dependence of single-shot all-optical switching in a GdFeCo film, which is susceptible to both the pulse width [S3] and starting temperature [S4].

To evaluate the pulse duration, we tested the ability of the optical pulses to achieve single-shot helicity-independent all-optical switching in a GdFeCo sample. Previous experiments have shown that this process depends not on $\lambda$ but rather on the pulse duration $\tau$ [S3]. If $\tau > \tau_c$ where $\tau_c$ is a cutoff duration, the pulse demagnetizes the optically-exposed area, whereas $\tau < \tau_c$ results in single-shot switching of magnetization. Moreover, $\tau_c$ is linearly dependent on the starting temperature $T_0$ so that as $T_0$ increases, $\tau_c$ decreases [S4]. This implies that for a fixed pulse duration, there is a cutoff starting temperature $T_{cutoff}$ above which the switching is disabled (see Fig. S4(a)).

Using these results, we can assess the duration of the optical pulses delivered by the OPA. A 20 nm-thick sample of $Gd_{23}(FeCo)_{77}$ is mounted on a resistive heater, and exposed to optical pulses of varying $\lambda$ as supplied by the OPA. Details of the experimental setup are otherwise the same as those given in Ref. [S3].

For a fixed $\lambda$, $T_0$ is steadily increased until single-shot all-optical switching is no longer possible, i.e., we identify $T_{cutoff}$. This method is repeated for varying $\lambda$, producing a dependence $T_{cutoff}(\lambda)$ as shown in Fig. S4(b). Next, by varying the duration of the 800 nm-pulses, we measure a state diagram for switching as a function of both τ and $T_0$. This is shown in Fig. S4(c). Linear fitting of the boundary between the two regions yields the dependence $\tau_c(T_0)$. Finally, by exploiting the known spectral indifference of $\tau_c$, we transform the cutoff starting temperature to the cutoff duration, thus obtaining the duration of the visible and near-infrared pulses as shown in Fig. S4(d).

**Supplemental Note 4: The spectral dependence of AO-HDS for other Pt/Co/Pt multilayers**

To obtain a more robust insight into the mechanism, we also investigated the spectral dependence of AO-HDS for other heterostructures. Here, we first studied a Pt (3 nm)/Co (0.8 nm)/Pt (3 nm) multilayer with the same procedures employed to investigate the Pt (3 nm)/Co (0.6 nm)/Pt (3 nm) stack. The magneto-optical properties of this heterostructure [Fig. S5 and Table S3] and AO-HDS [Fig. S6] were measured at $\lambda = 0.8$ μm, 0.7 μm, 0.6 μm, and 0.5 μm. Using these results, we evaluated the switching efficiency $\varepsilon$ [Fig. S7(a)] and effective fields of the inverse Faraday effect (IFE), $B_{\mathrm{eff}}^{\mathrm{IFE}}$, and the MCD effect, $B_{\mathrm{eff}}^{\mathrm{MCD}}$, [Fig. S7(b)] as a function of $\lambda$.

The $\varepsilon$ is ~0 % at $\lambda = 0.5$ μm but dramatically increases up to ~90 % at $\lambda = 0.8$ μm [Fig. S7(a)]. This striking increase of $\varepsilon$ is predominately attributed to the MCD effect, which increases by a factor of 7.4 in this spectral range [Fig. S6(b)]. On the other hand, the IFE non-monotonically changes in this spectral range [Fig. S6(b)]. These tendencies are commonly observed in the Pt (3 nm)/Co (0.6 nm)/Pt (3 nm) multilayer, showing that the conclusion of the main manuscript is robust even if the thickness of the ferromagnetic material changes.

We next investigated AO-HDS for a [Pt (2 nm)/Co (0.8 nm)]$_2$/Pt (2 nm) in the spectral range of $\lambda = 0.5 - 2.0$ μm. Figures S8 and S9 show typical snapshots of the AO-HDS and the switching efficiency $\varepsilon$ as evaluated by analyzing snapshots of AO-HDS for $\lambda = 0.5 - 2.0$ μm, respectively.

Increasing $\lambda$ leads to an increase in $\varepsilon$, which agrees with the cases of the other Pt/Co/Pt multilayers. The rising point of $\varepsilon$ in Fig. S9 shifts to longer $\lambda$ than for the other cases. Upon increasing the total thickness of the ferromagnetic stack, the size of magnetic domains (see Ref. [S5]) decreases, and the laser's intensity profile across the thickness becomes non-uniform. Hence, a larger MCD may be required to improve the $\varepsilon$.

TABLE S1. The refractive indices used for the transfer-matrix analysis. For the transfer-matrix method, we used refractive indices of Co, Pt, Ta, MgO, $Ta_2O_5$, and glass in Ref. S6-S11, respectively.

| Central Wavelength $\lambda$ (nm) | Refractive index $\tilde{n} = n + jk$ | | | | | |
|---|---|---|---|---|---|---|
| | $Ta_2O_5$ | MgO | Pt | Co | Ta | glass |
| 500 | 2.159 | 1.746 | $1.991 + j3.448$ | $2.025 + j3.720$ | $2.738 + j3.479$ | 1.462 |
| 600 | 2.128 | 1.734 | $2.268 + j3.967$ | $2.268 + j4.202$ | $2.007 + j4.032$ | 1.458 |
| 700 | 2.109 | 1.731 | $2.566 + j4.488$ | $2.566 + j4.480$ | $1.309 + j4.619$ | 1.453 |
| 800 | 2.096 | 1.728 | $2.839 + j4.950$ | $2.839 + j4.712$ | $1.117 + j3.526$ | 1.453 |
| 950 | 2.083 | 1.724 | $3.283 + j5.591$ | $3.283 + j4.998$ | $1.000 + j2.700$ | 1.451 |
| 1025 | 2.078 | 1.722 | $3.522 + j5.889$ | $3.522 + j5.128$ | $0.982 + j1.878$ | 1.450 |
| 1100 | 2.074 | 1.721 | $3.773 + j6.153$ | $3.773 + j5.258$ | $0.945 + j1.968$ | 1.449 |

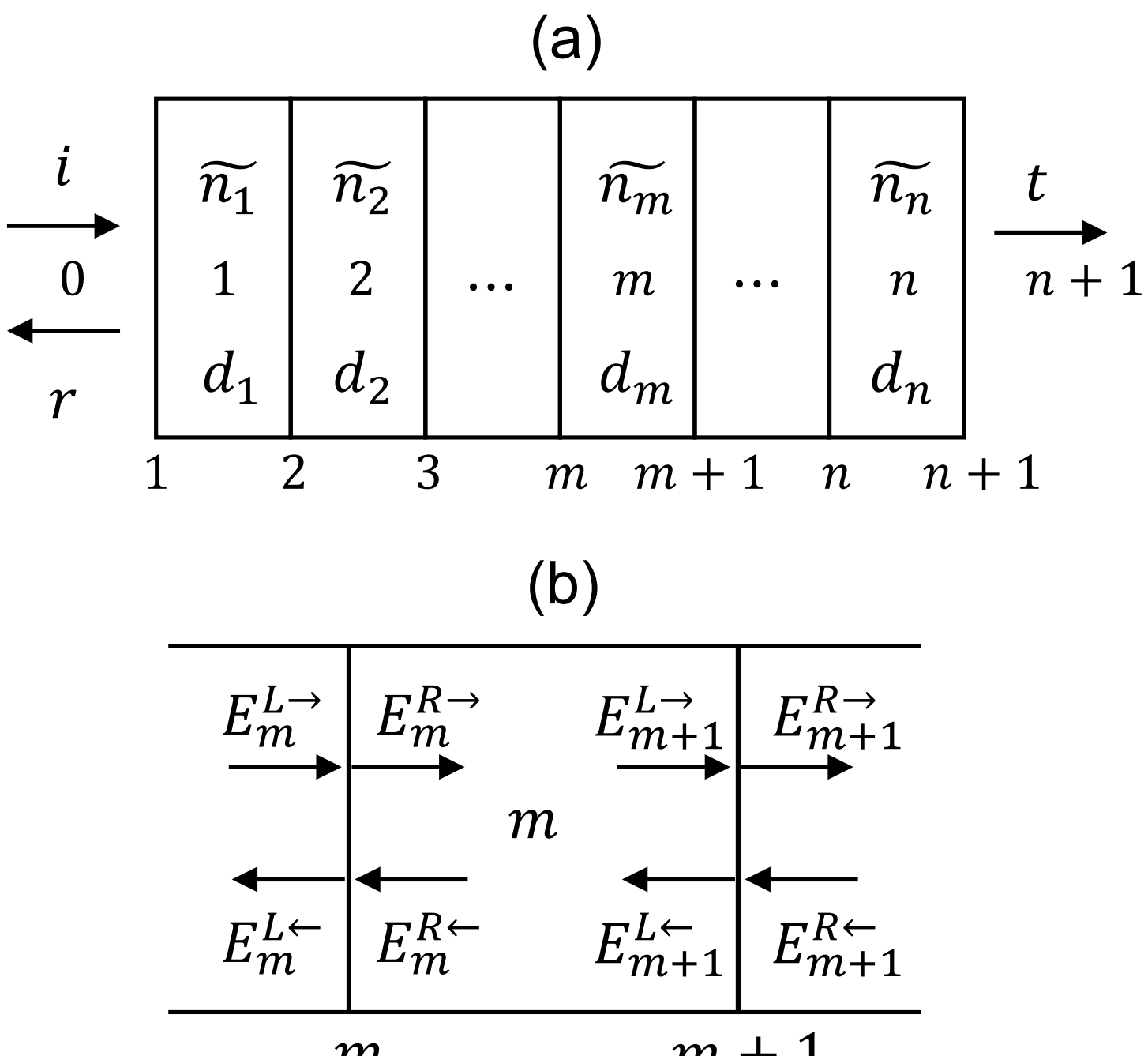

FIG S1. (a) A schematic of a thin multilayer with $n$ layers. (b) A schematic of the $m$-th layer with the light electric field amplitudes. The superscript letters and arrows denote the side of the interface and the light propagation direction, respectively. For instance, the notation $E_m^{L\rightarrow}$ corresponds to the electric field incident to the left side of the $m$-th interface.

TABLE S2. The magneto-optical parts of the refractive indices of the Co and Pt layers.

| Central Wavelength $\lambda$ (nm) | Refractive index ($M^{\uparrow}$) $\tilde{n} = n + jk$ | | | |
|---|---|---|---|---|
| | Co | | Pt | |
| | $\sigma^+$ | $\sigma^-$ | $\sigma^+$ | $\sigma^-$ |
| 500 | 2.042 + $j$3.680 | 2.009 + $j$3.760 | 1.987 + $j$3.408 | 1.954 + $j$3.488 |
| 600 | 2.064 + $j$4.132 | 2.589 + $j$4.272 | 2.281 + $j$3.897 | 2.230 + $j$4.037 |
| 700 | 3.210 + $j$4.365 | 3.142 + $j$4.595 | 2.574 + $j$4.373 | 2.506 + $j$4.603 |
| 800 | 3.671 + $j$4.529 | 3.565 + $j$4.895 | 2.892 + $j$4.767 | 2.786 + $j$5.133 |
| 950 | 4.058 + $j$4.635 | 3.896 + $j$5.342 | 3.364 + $j$5.247 | 3.202 + $j$5.935 |
| 1025 | 4.159 + $j$4.726 | 3.986 + $j$5.530 | 3.609 + $j$5.487 | 3.436 + $j$6.291 |
| 1100 | 4.277 + $j$4.749 | 4.059 + $j$5.767 | 3.882 + $j$5.644 | 3.665 + $j$6.662 |

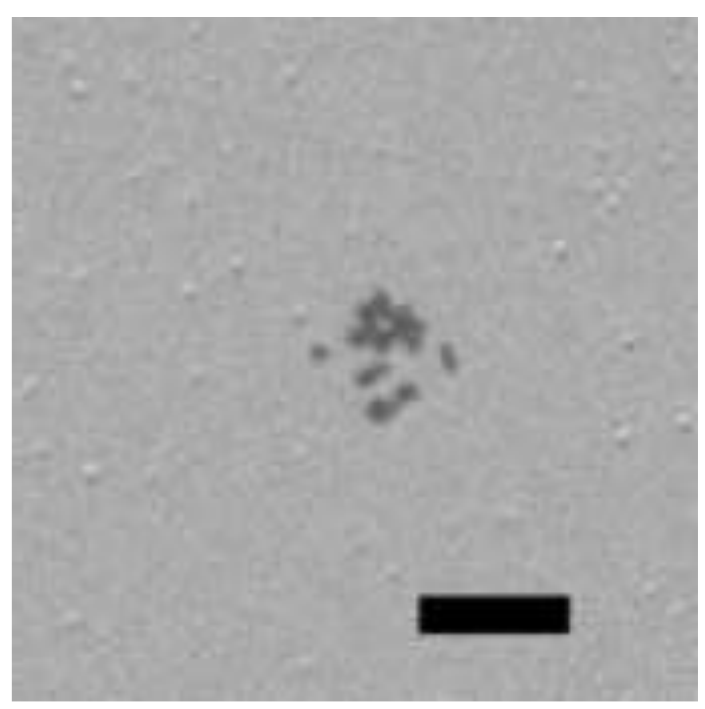

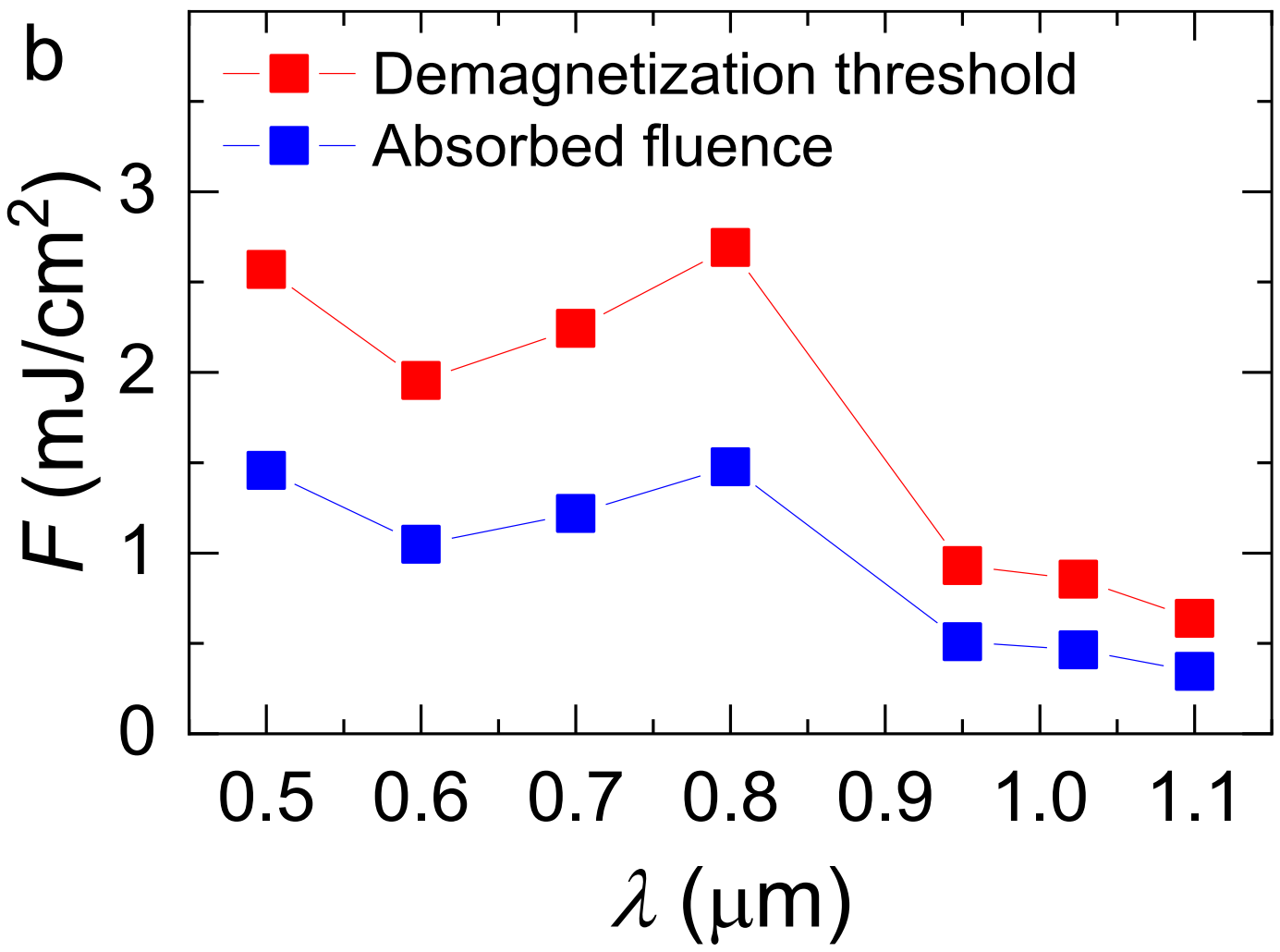

FIG S2. (a) A typical magneto-optical snapshot at the demagnetization threshold. The scale bar corresponds to 10 μm. (b) The spectral dependence of the demagnetization threshold. The absorbed fluences were calculated by considering the total light absorption by the multilayer.

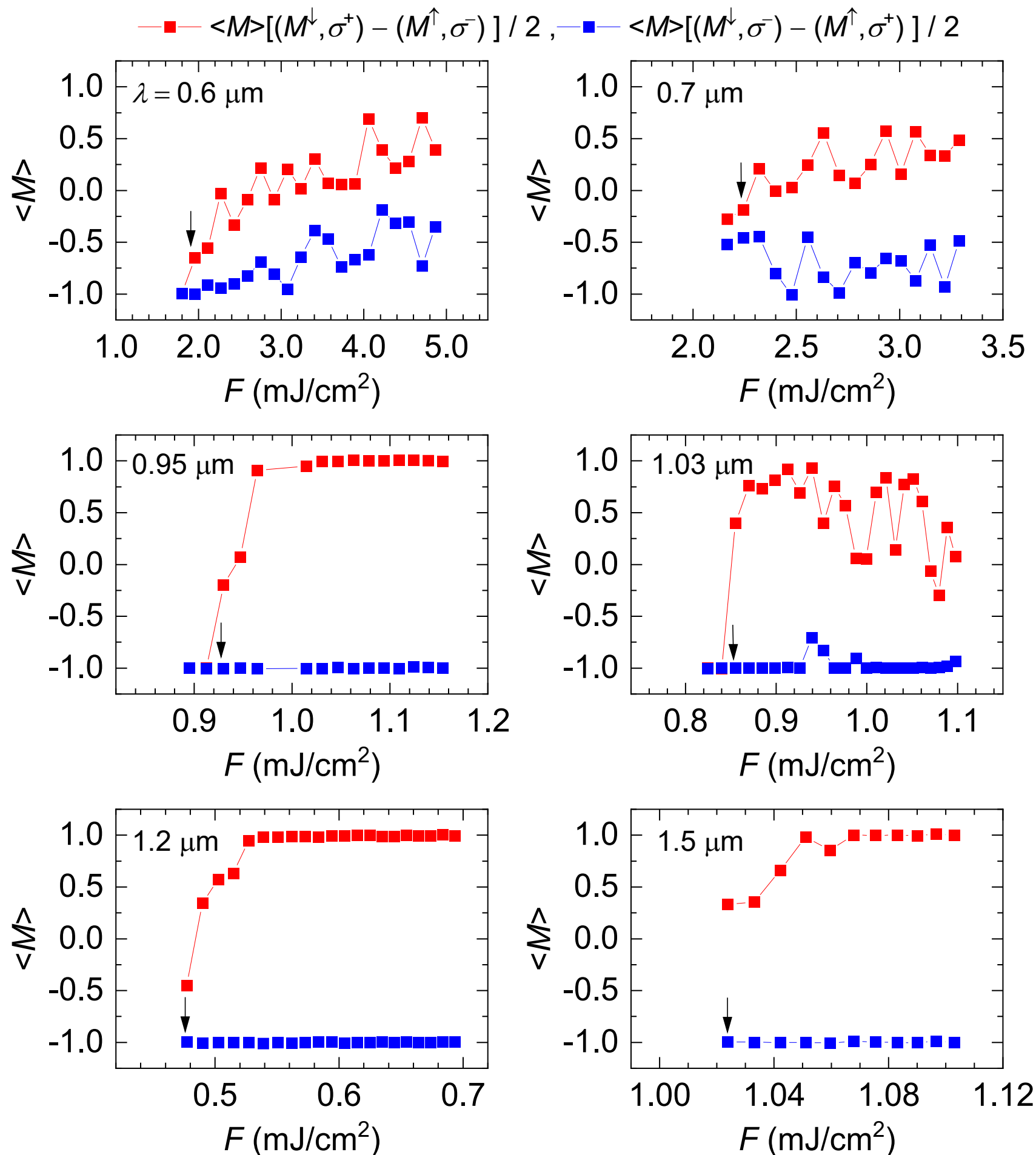

FIG. S3. Average net magnetization <*M*> as a function of Fluence *F* at $\lambda$ = 0.6 μm, 0.7 μm, 0.95 μm, 1.03 μm, 1.2 μm, and 1.5 μm for the Pt(3.0 nm)/Co(0.6 nm)/Pt(3.0 nm) multilayer stack. The black arrows indicate the demagnetization thresholds. We used the average value of the net magnetizations above the demagnetization threshold to calculate the switching efficiency shown in Fig 3(a).

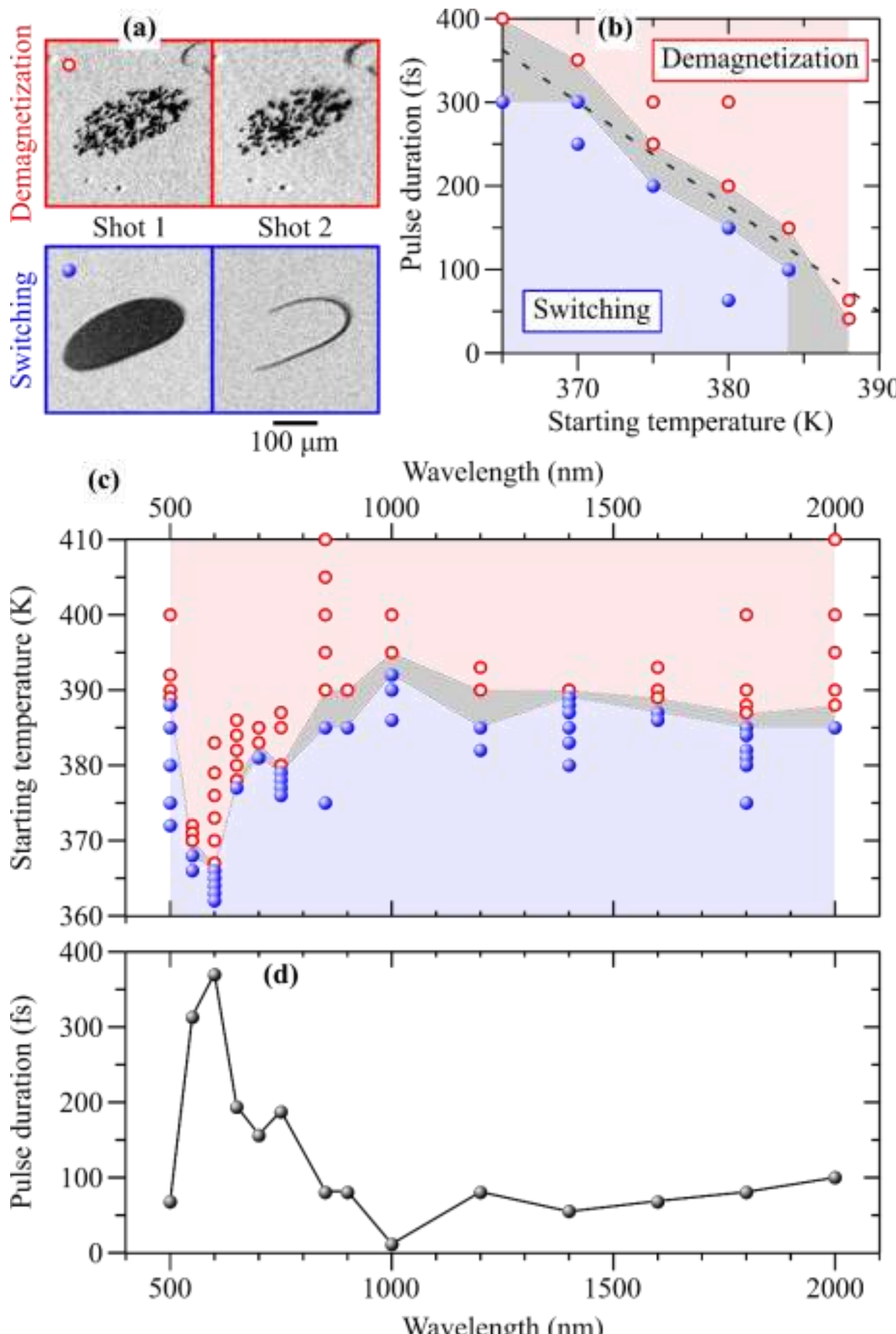

FIG. S4. (a) Typical magneto-optical images recorded after exposing GdFeCo to consecutive optical pulses at $\lambda = 1.8$ μm. The images in the top row were measured with a sample temperature $T_0 = 388$ K, whereas in the bottom row, $T_0 = 295$ K. (b) State diagram showing whether switching (blue area) or demagnetization (red area) is achieved by pulses at $\lambda = 0.8$ μm, with the indicated pulse duration and $T_0$. (c) State diagram showing whether switching is successfully obtained with different λ and $T_0$. (d) Estimated spectral dependence of the pulse duration delivered by the OPA, obtained using the results shown in panel (c) and the assumption that the cutoff temperature for switching is wavelength-independent.

TABLE S3. The magneto-optical parts of the refractive indices of the Co and Pt layers of the Pt (3 nm)/Co (0.8 nm)/Pt(3 nm) multilayer.

| Central Wavelength $\lambda$ (nm) | Refractive index ($M^{\uparrow}$) $\tilde{n} = n + jk$ | | | |
|---|---|---|---|---|
| | Co | | Pt | |
| | $\sigma^+$ | $\sigma^-$ | $\sigma^+$ | $\sigma^-$ |
| 500 | 2.042 + $j$3.680 | 2.009 + $j$3.760 | 1.987 + $j$3.408 | 1.954 + $j$3.488 |
| 600 | 2.064 + $j$4.132 | 2.589 + $j$4.272 | 2.281 + $j$3.897 | 2.230 + $j$4.037 |
| 700 | 3.210 + $j$4.365 | 3.142 + $j$4.595 | 2.574 + $j$4.373 | 2.506 + $j$4.603 |
| 800 | 3.671 + $j$4.529 | 3.565 + $j$4.895 | 2.892 + $j$4.767 | 2.786 + $j$5.133 |

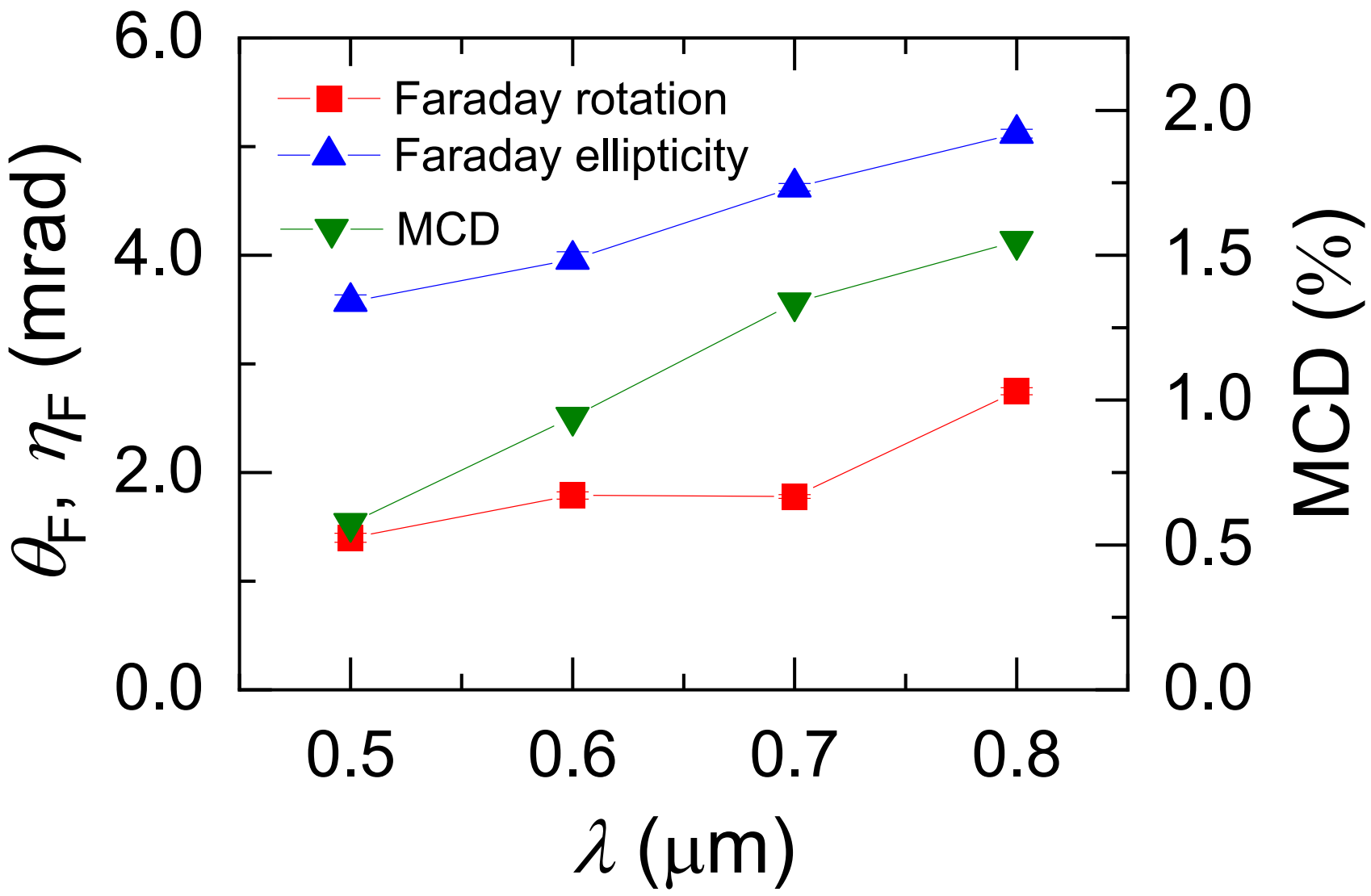

FIG. S5. The experimentally-measured spectral dependences of Faraday rotation $\theta_F$, ellipticity $\eta_F$ and magnetic circular dichroism MCD of a multilayered stack consisting of Ta (1 nm)/MgO (2 nm)/Pt (3 nm)/Co (0.8 nm)/Pt (3 nm)/Ta (4 nm) on a glass substrate.

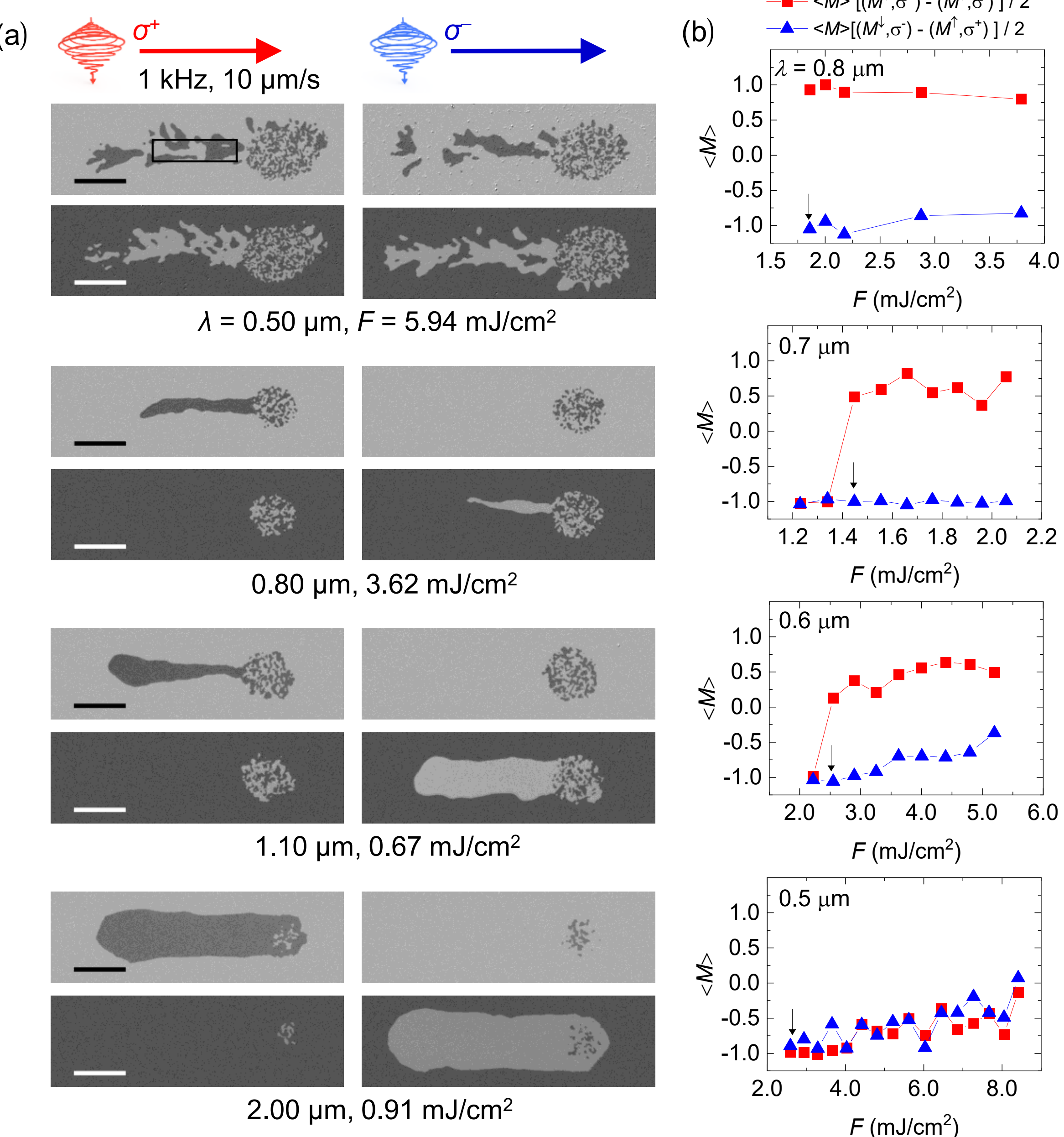

FIG. S6. (a) The spectral dependence of multi-pulse AO-HDS for the Pt (3 nm)/Co (0.8 nm)/Pt (3 nm) multilayer stack. The laser spot was swept at a speed of 10 μm/s. The focused $1/e^2$ radii were 47 μm, 61 μm, 46 μm, and 27 μm for $\lambda = 0.80$ μm, 0.70 μm, 0.60 μm, and 0.50 μm, respectively. The scale bar is 30 μm. (b) Average net magnetization <*M*> extracted from a section of the track versus the incident fluence *F*. A 20 μm × 80 μm rectangular area for integration is indicated in the snapshot. We determined <*M*> using the same method as used in Fig. 2(b) in the main text.

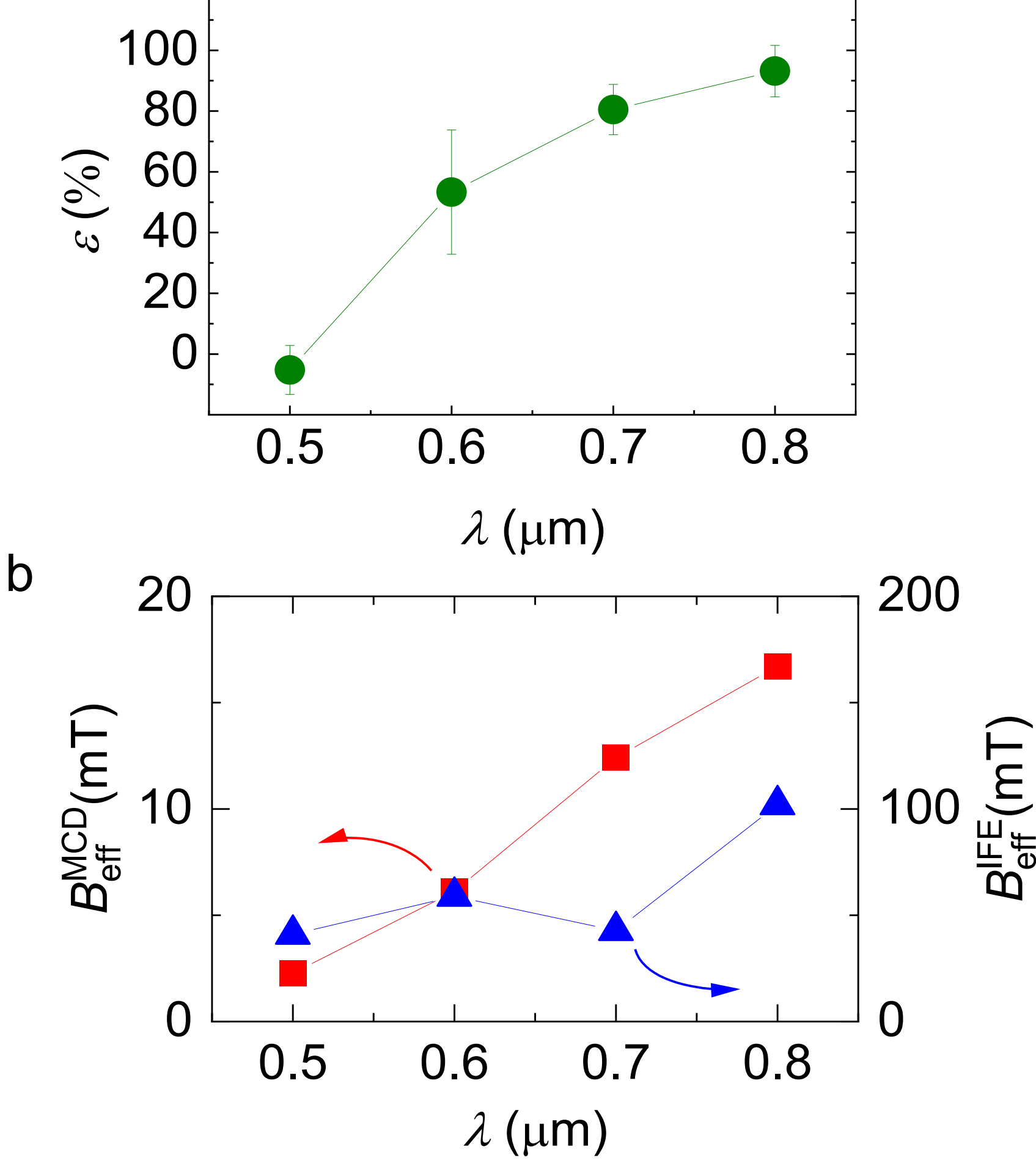

FIG. S7. (a) The spectral dependence of the switching efficiency $\varepsilon$ measured for the Pt (3 nm)/Co (0.8 nm)/Pt (2 nm) multilayer stack. (b) The spectral dependencies of the effective fields $B_{eff}$ associated with the inverse Faraday effect (blue triangles) and the MCD effect (red squares). When evaluating the $B_{eff}$ of the MCD effect, we used the magnetic parameters given in Ref [S12].

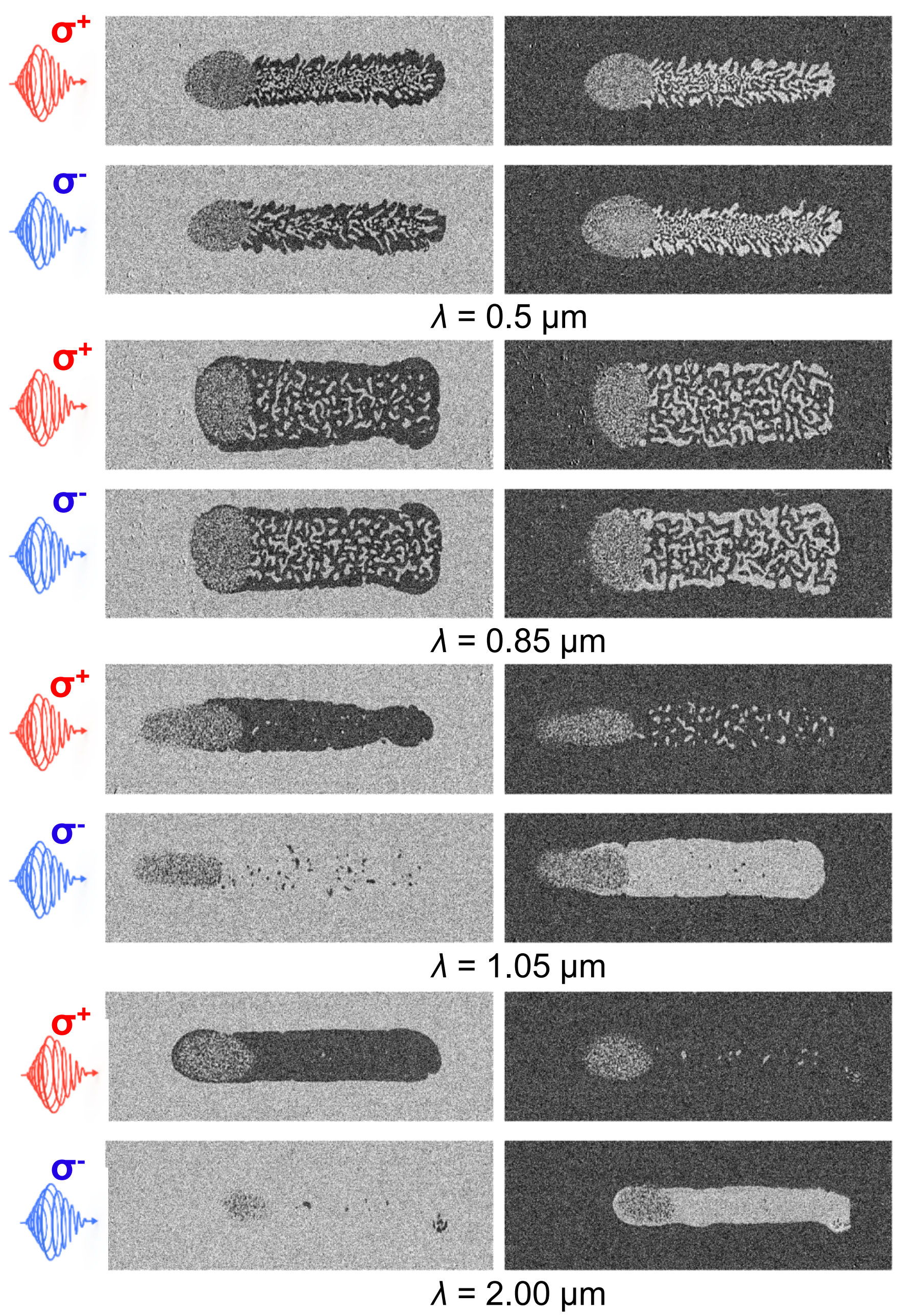

FIG. S8. The spectral dependence of multi-pulse AO-HDS for a multilayer consisting of Ta (1 nm)/MgO (2 nm)/[Pt (2 nm)/Co (0.8 nm)]$_2$/Pt (2 nm)/Ta (4 nm) grown on a glass substrate. The laser spot was swept across a distance of 250 μm at a speed of 10 μm/s. The size of the snapshots is 520 μm × 130 μm.

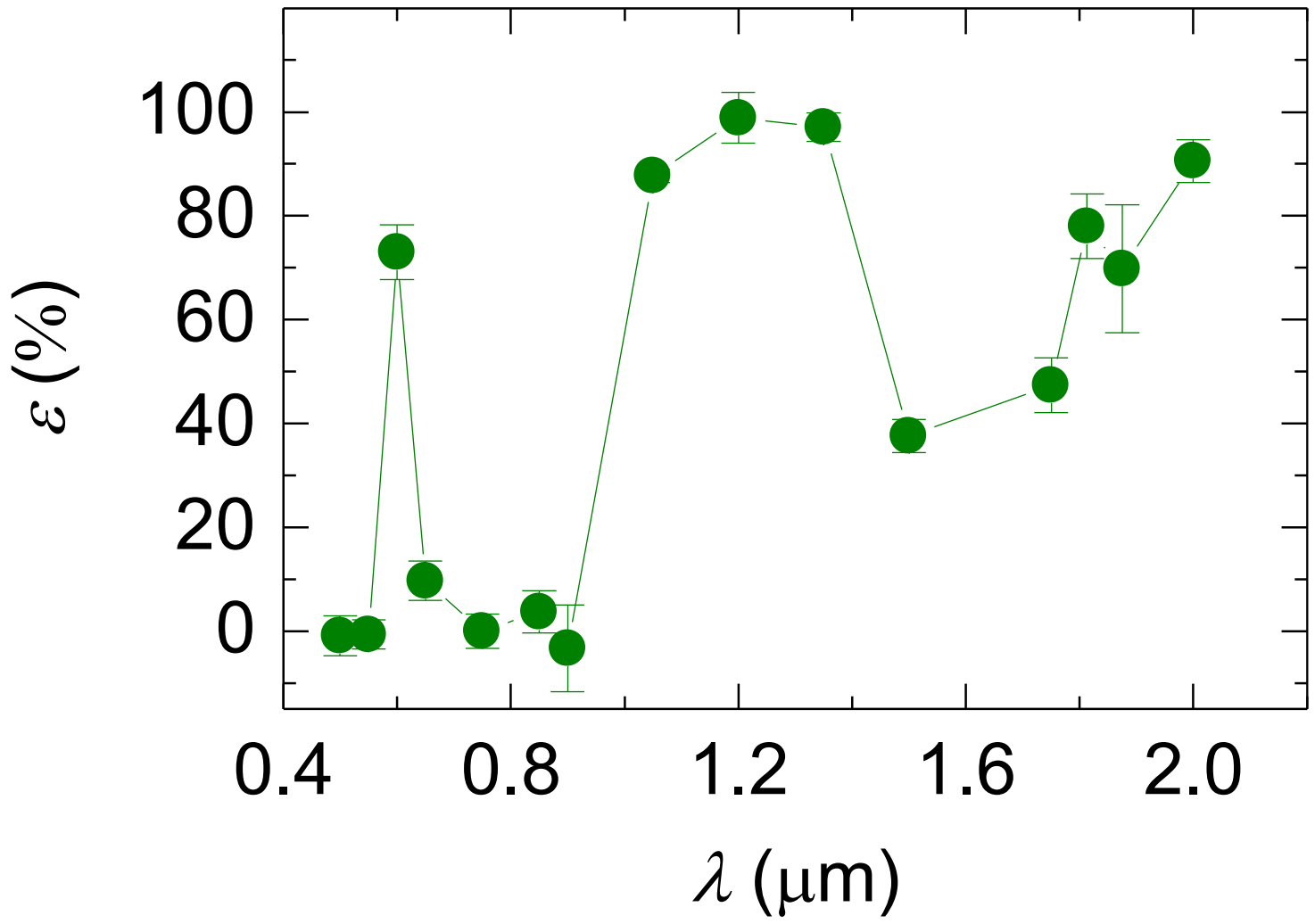

FIG. S9. (a) The spectral dependence of the switching efficiency $\varepsilon$ measured for the $[Pt\ (2\ nm)/Co\ (0.8\ nm)]_2/Pt\ (2\ nm)$ multilayer stack. For this evaluation, we averaged the intensities of 400 × 30 pixels, over which the laser spot was swept.